 \documentstyle[aps,prl,multicol,epsbox]{revtex}
\draft

\begin{document}

\title{Stripe Structures and the Berry-Phase Connection:\\
Concept of Geometric Energy}

\author{Yasutami Takada\thanks{To whom correspondence 
should be addressed. E-mail\ takada@issp.u-tokyo.ac.jp}
 and Takashi Hotta\thanks{
Present address: National High Magnetic Field Laboratory 
and Department of Physics, Florida State University,
Tallahassee, \qquad FL 32306, USA.}}

\address{Institute for Solid State Physics, University of Tokyo, 
  7-22-1 Roppongi, Minato-ku, Tokyo 106-8666, Japan}

\author{Hiroyasu Koizumi\thanks{Present address: 
Department of Chemistry and Biochemistry, 
University of Texas, Austin, TX 78712, USA.}}

\address{Faculty of Science, Himeji Institute of Technology,
Kanaji, Kamigori, Ako-gun, Hyogo 678-1297, Japan}

\date{\today}

\maketitle
\begin{abstract}
Electronic states of an $e_g$ electron are calculated in the 
system composed of two MnO$_6$ octahedra with the inclusion of 
the Berry phase acquired by parallel transport. 
Based on this calculation, a comment is made on the controversy 
between ``Wigner-crystal'' and ``paired-stripe'' models for the 
the insulating charge-ordered manganese oxides. 
\end{abstract}

\pacs{}

\begin{multicols}{2}
\narrowtext

\section{Introduction}

In quantum mechanics the Hamiltonian $H$ alone does not 
determine energy eigenstates of the system. 
We need information on the wavefunction $\varphi$ 
which includes the boundary condition in general and for the 
case of a many-electron system, antisymmetry due to the 
Fermi-Dirac statistics. 
The antisymmetry may be regarded as addition of the phase 
factor $e^{i\pi}$ to $\varphi$ in the interchange of any pair 
of electrons of the same spin. 
Note that this factor $e^{i\pi}$ brings about an important 
concept of ``the exchange interaction'' $J$ which is just 
the energy difference between singlet ($S=0$) and triplet 
($S=1$) states in an interacting two-electron system. 

Addition of the phase factor to $\varphi$ also occurs in the 
course of Berry-phase connection; the Berry phase is a geometric 
phase in $\varphi$ acquired by parallel transport.\cite{berry} 
On the analogy of $J$, we propose a new concept of 
``the geometric energy'' by considering an energy difference 
between the states characterized by different winding numbers 
$w$ (the Chern integers) associated with the topological 
invariant in parallel transport. 

In order to substantiate the above proposal, we calculate the 
geometric energy in this paper in the simplest possible system, 
namely, a system consisting of a single electron moving back and 
forth between two Jahn-Teller(JT) centers. 
In Sec. 2 we specify our system precisely by writing $H$ and 
a prescription for parallel transport. 
The lowest-energy state for each $w$ is given in Sec. 3 and 
Sec. 4 deals with our perspective into the stripe structures 
observed in La$_{1-x}$Ca$_{x}$MnO$_{3}$ with $x \geq 0.5$.
\cite{cheong}

\section{Two-Jahn-Teller-Center System and Parallel Transport}

Consider an $e_g$ electron in the system composed of 
two Mn$^{4+}$O$_6$ octahedra with one oxygen ion common 
to both octahedra to provide a transfer path with $t$ 
the transfer integral for the $e_g$ electron in $z$-direction. 
We assume that the spins of $t_{2g}$ electrons are aligned 
in one direction. 
Because of the strong Hund's rule coupling, spin of the 
$e_g$ electron also aligns in the same direction. 
In this sense, there are no spin degrees of freedom 
and we may write $H$ in second quantization as
\begin{eqnarray}
\label{eq:H}
  H \!&=& \!
   -t(b^{+}_1b_2+b^{+}_2b_1)
  +E_{\rm JT} \sum_{j=1,2} \Bigl [ 2{\bf d}_{j}^{+} 
  (q_{j}^{(2)} \tau_x \!+\! q_{j}^{(3)} \tau_z) {\bf d}_{j}
\nonumber \\
&&+{q_{j}^{(2)}}^2+{q_{j}^{(3)}}^2 \Bigr ]\,,
\end{eqnarray}
with the Pauli matrices, $\tau_{x}$ and $\tau_z$, and 
${\bf d}_{j}^{+} \equiv (a_{j}^{+}, b_{j}^{+})$, 
where $a_{j}^{+}$ and $b_{j}^{+}$ are, respectively, 
creation operators for electrons in $d_{x^2-y^2}$ and 
$d_{3z^2-r^2}$ orbitals at Mn site $j(=1\ {\rm or}\ 2)$, 
$q_{j}^{(2)}$ and $q_{j}^{(3)}$ are dimensionless vibrational 
variables representing, respectively, 
$(\delta X_{j}-\delta Y_{j})/\sqrt{2}$ and 
$(2\delta Z_{j}-\delta X_{j}-\delta Y_{j})/\sqrt{6}$ modes 
around site $j$, and $E_{\rm JT}$ is the static JT energy. 
Since we shall treat static distortions in the adiabatic 
approximation, the kinetic-energy term of ions is not 
considered in $H$.

The Berry-phase effect at site $j$ is best included 
by writing vibrational modes in polar coordinates as 
$q_{j}^{(2)}\!=\!q_{j} \sin \theta_{j}$ and 
$q_{j}^{(3)}\!=\!q_{j} \cos \theta_{j}$. 
Using $\theta_{j}$, we transform the operators 
$a_{j}$ and $b_{j}$ into ${\tilde a}_{j}$ and ${\tilde b}_{j}$ 
as ${\tilde a}_j\!=\!e^{i\theta_j/2} [a_j \cos (\theta_{j}/2) 
\!+\! b_j \sin (\theta_{j}/2)]$ and 
${\tilde b}_j\!=\!e^{i\theta_j/2} [-a_j \sin (\theta_{j}/2) 
\!+\!b_j \cos (\theta_{j}/2)]$. 
In this representation, the second term (the site-energy term) 
in (\ref{eq:H}) is diagonalized. 
The molecular Aharonov-Bohm effect manifests itself 
in the phase factor $e^{i \theta_{j}/2}$ at each site $j$ 
at which topology of the singularity is understood 
as a double-plane feature. 

The transfer between the two sites connects $\theta_{1}$ with 
$\theta_{2}$. 
We make this connection by parallel transport, by which we mean 
that if we start with a state representing an electron at site 1, 
$\varphi_1$, we make the state evolve from $\varphi_1$ 
in the direction orthogonal to it, namely, along the direction 
of a state representing an electron at site 2, $\varphi_2$. 
This parallel transport indicates that an $e_g$ electron 
picks up only a constant phase $\theta_{0}$ 
by a trip from site 1 to site 2, namely, 
$\theta_{2}\!=\!\theta_{1}+\theta_{0}$. 
A further trip in the same direction from site 2 results in 
the total phase as $\theta_{2}+\theta_{0}\! 
=\!\theta_{1}+2\theta_{0}$. 
Now in general, a two-site system is equivalent to a 
corresponding periodic lattice system with periodicity two. 
Thus the phase $\theta_{1}+2\theta_{0}$ should represent the 
same physics at site 1 at which the phase is $\theta_{1}$. 
This reasoning leads us to the conclusion that 
$2\theta_{0}$ should be equal to $2\pi w$ with an integer $w$. 
If we make a more general argument on topology,\cite{hotta} 
it turns out that $w$ is nothing but the winding number, 
a topologically conserved quantity which is useful to specify 
an eigenstate of the system. 
Note that $w$ describes a geometric structure as to 
how the double-plane structure at each site is interconnected. 

\section{Geometric Energy}

\begin{figure}[h]
\center
\epsfile{file=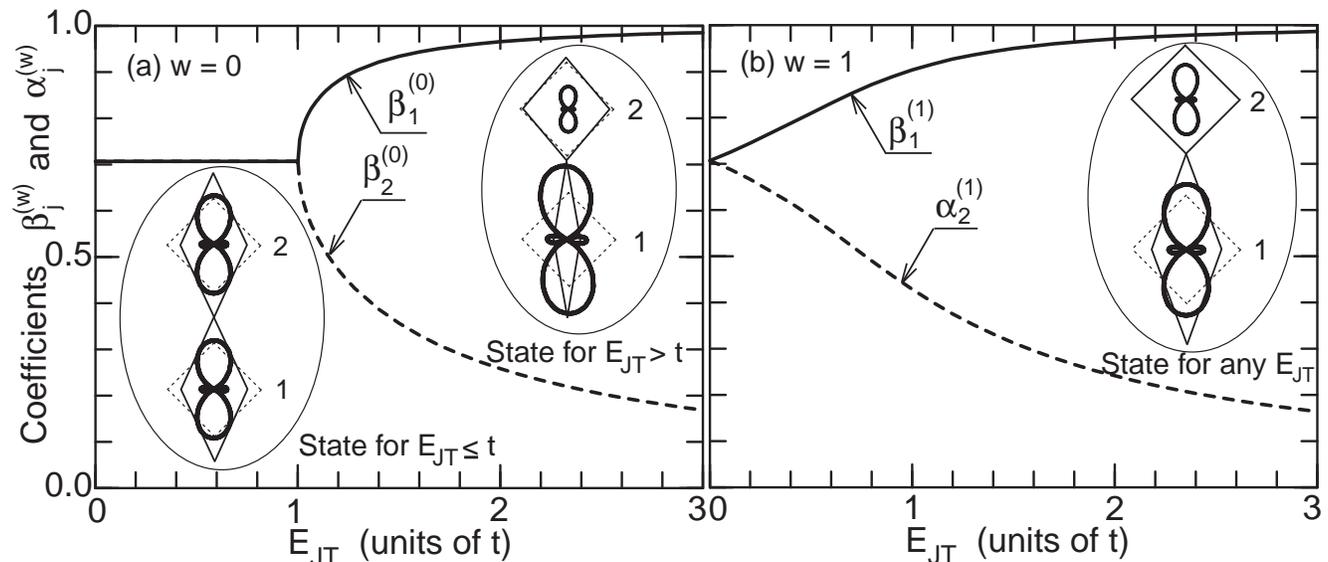,scale=0.9}
\caption{Coefficients to specify the lowest single-electron 
wavefunctions plotted as a function of $E_{\rm JT}$ 
for (a) $w=0$ and (b) $w=1$.} 
\label{fig1}
\end{figure}

Let us calculate the lowest single-electron energy for each 
$w(=0\ {\rm or}\ 1)$, $E_0^{(w)}$. 
We adopt a variational procedure by considering a 
wavefunction $|\Psi_0^{(w)} \rangle$ as 
\begin{eqnarray}
  |\Psi_0^{(w)} \rangle \!&=&\! \sum_{j=1,2} e^{i\theta_j/2}
 (\alpha_j^{(w)}{\tilde a}_{j}^{+}
  + \beta_j^{(w)}{\tilde b}_{j}^{+})
|{\rm vac} \rangle 
\nonumber \\
&&\otimes |q_1=\bar{q}_1,q_2=\bar{q}_2 \rangle\,,
\end{eqnarray}
where $\alpha_j^{(w)}$ and $\beta_j^{(w)}$ are real numbers, 
$|{\rm vac} \rangle$ is the vacuum for electron operators, 
and $|q_1=\bar{q}_1,q_2=\bar{q}_2 \rangle$ is the phonon 
wavefunction representing the $\delta$-function-like distribution 
of $q_j$ around $\bar{q}_j$.
We have optimized all the parameters involved in the problem 
except for $\theta_2$ (which is related to $\theta_1$ through 
$\theta_1+\pi w$) in order to make 
$\langle \Psi_0^{(w)}|H|\Psi_0^{(w)} \rangle$ minimum 
under the condition of 
$\langle \Psi_0^{(w)}|\Psi_0^{(w)} \rangle = 1$ for 
given $t$ and $E_{\rm JT}$. 
We have found that the lowest energy is obtained at $\theta_1=0$ 
irrespective of $w$. 
This is the condition for the $e_g$-electron orbital 
to polarize in the transfer direction. 
(If we start with a different configuration, say two sites 
in $x$-direction, we obtain completely the same physical results 
with the interchange of the role of $z$-axis with that of $x$.) 
However, $|\Psi_0^{(w)} \rangle$ is different for different $w$ 
in an interesting way.
For $w=0$, we obtain
\begin{eqnarray}
|\Psi_0^{(0)} \rangle 
&=& (\beta_1^{(0)}b_1^{+}+\beta_2^{(0)}b_2^{+})
|{\rm vac} \rangle 
\nonumber \\
&&\otimes 
|q_1={\beta_1^{(0)}}^2,q_2={\beta_2^{(0)}}^2 \rangle\,,
\end{eqnarray} 
with the coefficients $\beta_j^{(0)}$ shown in Fig.~\ref{fig1}(a) 
as a function of $E_{\rm JT}$. 
As long as $E_{\rm JT}$ is smaller than $t$, symmetry exists 
between the sites, but for larger $E_{\rm JT}$ the site-symmetry 
is broken; an $e_g$ electron tends to localize in site 1. 
(Of course, the state with the interchange of the sites is 
possible, but there is no overlap between these two states 
because of the orthogonality of phonon wavefunctions.) 

For $w=1$, on the other hand, we obtain
\begin{eqnarray}
|\Psi_0^{(1)} \rangle 
&=& (\beta_1^{(1)}b_1^{+}+\alpha_2^{(1)}b_2^{+})
|{\rm vac} \rangle 
\nonumber \\
&&\otimes 
|q_1={\beta_1^{(1)}}^2,q_2=0 \rangle\,,
\end{eqnarray} 
with the coefficients $\beta_1^{(1)}$ and $\alpha_2^{(1)}$ 
shown in Fig.~\ref{fig1}(b). 
In this case, the site-symmetry is broken even 
for an infinitesimally small positive value for $E_{\rm JT}$ 
due to the fact that the lattice at site 2 never deforms, 
because with this winding number, an $e_g$ electron at site 1 
in the lower adiabatic potential plane is parallel-transported 
to the upper adiabatic potential plane at site 2. 
In this sense, $w=1$ describes ``the inter-potential-plane 
connection'', while $w=0$ ``the intra-potential-plane 
connection''.

\begin{figure}[h]
\center
\epsfile{file=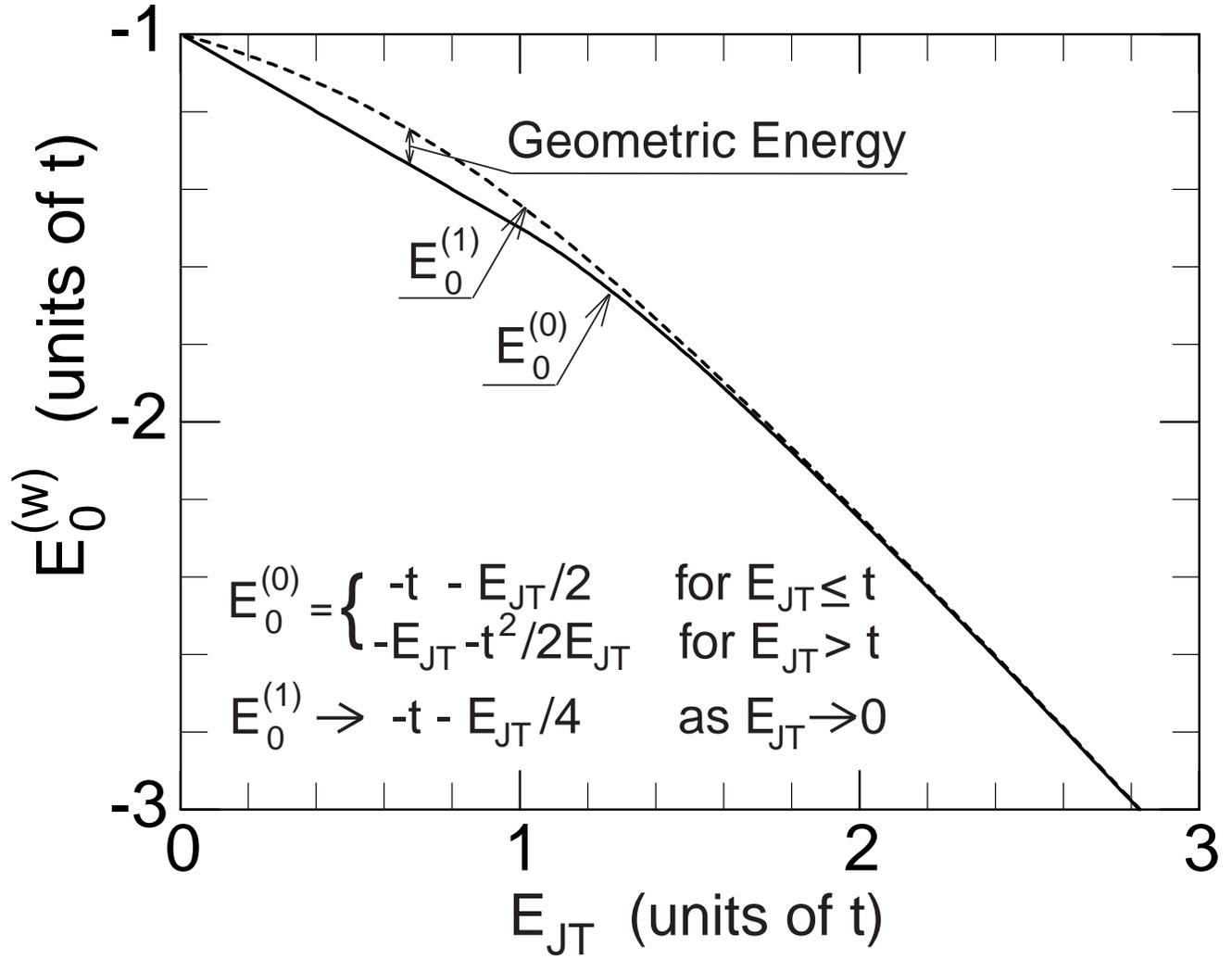,scale=0.9}
\caption{Lowest single-electron energies 
as a function of $E_{\rm JT}$ for $w=0$ and $1$.} 
\label{fig2}
\end{figure}

We plot the corresponding energies $E_0^{(w)}$ as a function 
of $E_{\rm JT}$ in Fig.~\ref{fig2} in which we give an analytic 
expression for $E_0^{(0)}$. 
(We can give $E_0^{(1)}$ only numerically.)
The term $-t^2/2E_{\rm JT}$ clearly indicates the hopping nature 
of an electron localized by the JT stabilization energy 
$-E_{\rm JT}$ for $E_{\rm JT}> t$. 
We find that $E_0^{(0)}$ is always lower than $E_0^{(1)}$. 
The geometric energy, $E_0^{(1)}-E_0^{(0)}$, is negligibly small 
in both large- and small-$E_{\rm JT}$ regions, but it becomes 
as large as $0.1t$ for $E_{\rm JT} \approx t$. 

Quite a similar situation occurs in the $J$-problem in a 
two-electron system in which the conserved quantity is the 
total spin $S$; the energy of the singlet state $E_0^{S=0}$ 
is always lower than that of the triplet one $E_0^{S=1}$. 
Note, however, that application of external magnetic fields 
can compensate the energy difference to stabilize the triplet 
state. 
Analogously, we may stabilize $|\Psi_0^{(1)} \rangle$ 
by compensating the geometric energy with application of 
external stresses to the system in $z$-direction, because 
the state $|\Psi_0^{(1)} \rangle$ is less distorted in the 
direction than $|\Psi_0^{(0)} \rangle$. 

\section{Discussion on Paired-Stripe Structures}

Considerable accumulation of experimental data has been made 
as for the crystal structures of the insulating 
La$_{1-x}$Ca$_{x}$MnO$_{3}$ with $x \geq 0.5$. 
Everyone agrees the existence of the charge and orbital 
ordering in these compounds, but a controversy continues about 
their detailed crystallographic superstructures; 
some claim the ``paired-stripe'' model\cite{cheong} and 
others the ``Wigner-crystal'' model.\cite{radaelli}

In order to shed light on the argument, we have extended 
our calculation from the two-JT-center problem to a 
two-dimensional lattice system.\cite{hotta} 
The key idea is to identify a zigzag conducting path of 
$e_g$ electrons along which we define the winding number $w$.
We have found that the state corresponding to the 
Wigner-crystal model provides the lower energy than 
the paired-stripe state, though the energy difference is 
quite small. 
This result is obtained without any consideration of the 
long-range Coulomb interaction $V$. 
We have included the effect of $V$ recently and found 
that further stabilization is obtained for 
the Wigner-crystal state over the paired-stripe one. 
Thus, in terms of energies, it seems to be certain that 
the former state is favorable. 

However, we have investigated the electronic state of 
$e_g$ electrons in the observed paired-stripe structure, 
evaluated the corresponding winding number $w$, and 
found an interesting relation between $w$ and 
the $e_g$-electron density specified by $x$ as\cite{hotta} 
\begin{equation}
w = {x \over 1-x} = 
{{\rm Number~of~Mn}^{4+}~{\rm ions} \over 
{\rm Number~of~Mn}^{3+}~{\rm ions}}\,.
\end{equation} 
Since $w$ is an integer, some specific values 
exist for $x [=w/(1+w)]$ such as $1/2$, $2/3$, $3/4$, etc.
This reminds us of the experimentally observed special 
values of $x$ and the related lever rule.\cite{cheong}
Thus, if the crystal is placed in an ``environment'' 
to favor the state with this winding number $w$, 
we can expect that the paired-stripe state may be observed. 
At present, we do not know precisely how we can define the 
``environment'' and this may be an important issue 
in the future not only from a viewpoint to clarify 
the crystal structures of La$_{1-x}$Ca$_{x}$MnO$_{3}$ 
but also from a fundamental aspect as to the identification 
of a physical variable to control the winding number 
or the geometric energy.

\acknowledgments

YT is supported by the Mitsubishi Foundation as well as
the Grand-in-Aid for Scientific Research (C) 
from the Monbusho of Japan.


\end{multicols}

\widetext

\begin{references}

\bibitem{berry}
M. V. Berry, Proc. R. Soc. {\bf A392}, 45 (1984);
J. Anandan, Nature {\bf 360}, 307 (1992).

\bibitem{cheong}
S. Mori, C. H. Chen, and S-W. Cheong, 
Nature {\bf 392}, 473 (1998); 
Phys. Rev. Lett. {\bf 81}, 3972 (1998). 

\bibitem{hotta}
T. Hotta, Y. Takada, and H. Koizumi, 
Int. J. Mod. Phys. B {\bf 12}, 3437 (1998).

\bibitem{radaelli}
P. G. Radaelli, D. E. Cox, M. Marezio, and S-W. Cheong, 
Phys. Rev. {\bf B55}, 3015 (1997);
P. G. Radaelli, D. E. Cox, L. Capogna, S-W. Cheong, and 
M. Marezio, cond-mat/9812366.


\end{references}
\end{document}